\documentclass[conference]{IEEEtran}
\IEEEoverridecommandlockouts

\usepackage{cite}
\usepackage{amsmath,amssymb,amsfonts}
\usepackage{graphicx}
\usepackage{textcomp}
\usepackage{xcolor}
\usepackage{booktabs}
\usepackage{url}
\usepackage{microtype}
\usepackage{listings}
\usepackage{tikz}
\usetikzlibrary{shapes.geometric, arrows.meta, positioning, fit, backgrounds}

\def\BibTeX{{\rm B\kern-.05em{\sc i\kern-.025em b}\kern-.08em
    T\kern-.1667em\lower.7ex\hbox{E}\kern-.125emX}}

\begin{document}

\title{Zero-Instrumentation Dependency Discovery for\\
Guided Microservice Migration Using eBPF}

 \author{\IEEEauthorblockN{Eshan Trivedi\textsuperscript{1} and Chandrahasa Pranava\textsuperscript{2}}
  \IEEEauthorblockA{\textsuperscript{1}Independent researcher, San Francisco, USA, eshanntrivedi@gmail.com\\
  \textsuperscript{2}Independent researcher, Dubai, UAE, vvscpranava@gmail.com}}

\maketitle

\begin{abstract}
Migrating microservices across virtual machines (VMs) without knowledge of
their runtime communication patterns risks creating cross-VM hotspots and
latency spikes that are difficult to predict from static analysis alone.
We use extended Berkeley Packet Filter (eBPF) kernel-level network tracing to
automatically discover inter-service dependencies at runtime, with no
application instrumentation, and use the resulting dependency graph to produce
a traffic-aware migration plan ranked by return on investment (ROI).
A two-pass process-identifier (PID) to port correlation algorithm recovers the
identity of all 20
services in a shared-runtime testbed where processes are otherwise
indistinguishable, matching the known ground-truth topology.
The system discovers 32 dependency edges from 13,615 network events captured in
three minutes, and applies spectral graph clustering with Kernighan-Lin refinement
to partition services into VM-coherent groups.
In simulation over the discovered graph, our ROI-ranked migration order reduces
cumulative cross-VM traffic exposure during the migration window by 27\%
relative to alphabetical ordering, a deterministic proxy for arbitrary
dependency-blind ordering.
Collection overhead is mixed: in a controlled A/B test at near-saturation load
on a host with two virtual CPUs (vCPUs), throughput fell by only 4.4\%, but
median (p50) latency rose by 383\% and 99th-percentile (p99) latency by
1,050\%.
We therefore recommend running captures off-peak or on dedicated sampling nodes
rather than under production saturation.
All results are from a single 20-service testbed that we authored; we make no
claim about behavior on production dependency graphs.
\end{abstract}


\section{Introduction}

Cloud operators routinely migrate microservices across virtual machines to
consolidate costs, satisfy service-level agreement (SLA) requirements, or
respond to hardware
failures~\cite{pahl2016microservices}.
The dominant practice is to move services one at a time in an order determined
by team ownership, alphabetical convenience, or rough intuition about which
services are ``less important.''
None of these approaches account for the runtime communication graph: which
service calls which, how often, and how much traffic crosses VM boundaries at
each step of the migration.

Moving a high-fan-in hub service early immediately exposes all of its callers
to cross-VM network latency for the rest of the migration window. Moving leaf
services last wastes the chance to co-locate them with their dependencies at
low cost. In simulation over our 20-service testbed graph, the worst-case naive
ordering sustains cross-VM traffic above 50\% of total flow volume for seven of
eleven migration steps; our approach keeps the same peak below 48\% and reduces
cumulative exposure by 27\%. We did not execute the migrations themselves.

The obstacle to doing better is that runtime dependency graphs are largely
invisible. Service meshes such as Istio~\cite{istio} and Linkerd~\cite{linkerd}
can expose this information, but they require sidecar injection, mutual
Transport Layer Security (TLS)
configuration, and ongoing operational overhead that many teams defer or avoid.
Static analysis of configuration files misses dynamic behavior. Distributed
tracing systems such as Jaeger~\cite{jaeger} require application-level
instrumentation.

We take a different approach: observe the kernel's Transmission Control
Protocol (TCP) layer directly using the extended Berkeley Packet Filter
(eBPF)~\cite{gregg2019bpf}, where every \texttt{connect} and \texttt{accept}
syscall is visible regardless of the application stack. This requires no code
changes, no sidecars, and no restart of running services. The probe attaches
and detaches at the operator's discretion.

The technical challenge is attribution: when all microservices run as
indistinguishable \texttt{python3} processes under the same OS user (a common
pattern in containerized or VM-hosted deployments), eBPF events carry no
service identity. We solve this with a two-pass PID-port correlation algorithm
that infers service identity from the ports each process listens on, then
attributes outbound connections accordingly.

This paper makes three contributions:

\begin{enumerate}
\item \textbf{PID-port correlation}: A two-pass algorithm that resolves service
identity in shared-runtime environments using only kernel-observable events,
recovering the correct identity of all 20 services in our testbed as measured
against the known ground-truth topology (\S\ref{sec:pid}).

\item \textbf{Spectral clustering with Kernighan-Lin (KL) refinement}:
Application of spectral graph clustering to the eBPF-derived dependency graph,
augmented with KL bisection refinement that directly minimizes the inter-VM
communication cut rather than Euclidean embedding distance (\S\ref{sec:cluster}).
Both algorithms are long established; we claim no algorithmic novelty here, only
their combination and their application to an automatically derived graph.

\item \textbf{ROI-ranked migration ordering}: A per-service ROI score that
balances traffic consolidation benefit against dependency-depth risk and
load variance, producing a phased migration plan evaluated against naive
baselines in simulation (\S\ref{sec:roi}).
\end{enumerate}

We report these as results on a single synthetic testbed. \S\ref{sec:limits}
states plainly what that does and does not establish.

\section{Background}

\subsection{eBPF and Network Tracing}

Extended Berkeley Packet Filter (eBPF) allows verified programs to run inside
the Linux kernel in response to events such as system calls, tracepoints, and
kprobes, without modifying kernel source or loading kernel
modules~\cite{gregg2019bpf}.
For network dependency discovery, the relevant hooks are
\texttt{tcp\_v4\_connect} (outbound TCP connection attempts) and
\texttt{inet\_csk\_accept} (accepted incoming connections).
Each event yields the source/destination Internet Protocol (IP) address and
port tuple, the process thread-group identifier (TGID, the kernel's identifier
for a process), and a nanosecond timestamp, which is enough to reconstruct
service-to-service communication graphs at any desired granularity.

The eBPF verifier enforces termination and memory safety, making probes safe to
attach to production kernels. The probe fires once per TCP connection
establishment rather than once per packet, which bounds the event rate even for
high-bandwidth services holding persistent connections. Bounding the event rate
is not the same as bounding cost, however: \S\ref{sec:overhead} shows that
draining those events competes for CPU with the application itself on a
resource-constrained host under load.

\subsection{The Microservice Migration Problem}

Given $n$ services initially co-located on a source VM, and a target topology
of $k$ VMs, the migration planning problem has two subproblems:
(1)~\emph{partition}: assign each service to a target VM to minimize
cross-VM traffic, and (2)~\emph{ordering}: determine the sequence in which
to move services to minimize cross-VM traffic \emph{during} the migration
window, not just at its end.

Both subproblems are NP-hard in general~\cite{andreev2004balanced}.
Spectral methods provide effective polynomial-time approximations for the
partition problem~\cite{ng2001spectral}. The ordering problem appears to have
received considerably less attention than the partition problem, and we did not
find prior work treating it as a dependency-aware optimization target. We note
that our survey is not exhaustive, and that commercial migration tooling is not
documented in enough detail to rule out similar heuristics being used in
practice.

\section{Related Work}

\textbf{Service dependency discovery.}
Dependency maps have been inferred from distributed tracing systems such as
Dapper~\cite{sigelman2010dapper} and Jaeger~\cite{jaeger}.
These approaches require instrumentation at the application layer, whether
through modified shared libraries or explicit span annotation. eBPF-based
dependency discovery has been demonstrated in the context of security
monitoring~\cite{falco} and performance profiling~\cite{gregg2019bpf}, but
not, to our knowledge, for migration planning.

\textbf{Service mesh observability.}
Istio~\cite{istio} and Linkerd~\cite{linkerd} expose rich service graphs via
their data planes, but require sidecar injection and place a proxy on the data
path, which consumes CPU per proxy and adds per-request latency. Istio publishes
per-version benchmark figures for this cost~\cite{istio-perf}; the magnitude
depends on release, configuration, and traffic pattern, so we do not quote a
single number.
Our approach is orthogonal: it works at the kernel layer, requires no
application changes, and can be applied to legacy or third-party services that
cannot be modified.

\textbf{VM and container migration.}
Live VM migration has been studied extensively~\cite{clark2005live,
nelson2005fast}, as has container placement~\cite{burns2016borg}.
These works focus on minimizing downtime during the move of a single service;
they do not address the multi-service ordering problem.
Commercial tools such as AWS Migration Hub and Azure Migrate operate on
inventory and cost data rather than runtime traffic graphs.

\textbf{Graph partitioning for service placement.}
Spectral partitioning is a standard technique for balanced graph
bisection~\cite{ng2001spectral, shi2000normalized}.
Prior work has applied it to data-center network
topology~\cite{mysore2009portland} and microservice
placement~\cite{luo2021characterizing}.
We extend this line of work by (a) deriving the input graph automatically from
eBPF telemetry and (b) adding KL refinement to directly optimize the cut metric
relevant to migration cost.

\section{System Design}
\label{sec:design}

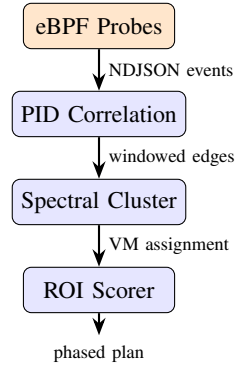
\begin{figure}[t]
\centering
\begin{tikzpicture}[
  box/.style={rectangle, rounded corners=3pt, draw, fill=blue!10,
              minimum width=2.0cm, minimum height=0.6cm,
              font=\small, text centered},
  arr/.style={-Stealth, thick},
  node distance=0.55cm and 0.3cm
]
\node[box, fill=orange!20] (ebpf)  {eBPF Probes};
\node[box, below=of ebpf]  (parse) {PID Correlation};
\node[box, below=of parse] (clust) {Spectral Cluster};
\node[box, below=of clust] (roi)   {ROI Scorer};

\draw[arr] (ebpf)  -- node[right, font=\scriptsize]{NDJSON events} (parse);
\draw[arr] (parse) -- node[right, font=\scriptsize]{windowed edges} (clust);
\draw[arr] (clust) -- node[right, font=\scriptsize]{VM assignment}  (roi);
\draw[arr] (roi)   -- +(0,-0.6) node[below, font=\scriptsize]{phased plan};
\end{tikzpicture}
\caption{Four-phase pipeline. Phases are decoupled; each writes a file consumed
by the next, enabling incremental reruns.}
\label{fig:pipeline}
\end{figure}

The system comprises four decoupled phases (Fig.~\ref{fig:pipeline}):

\textbf{Phase 1: eBPF capture.}
A lightweight nettrace daemon attaches kprobes to \texttt{tcp\_v4\_connect}
and \texttt{inet\_csk\_accept}. Each event is emitted as a JSON record
containing timestamp (nanosecond), source/destination port, direction flag, and
process TGID. The daemon writes append-only newline-delimited JSON (NDJSON); it
can be attached and
detached without restarting any service.

\textbf{Phase 2: PID-port correlation and windowing.}
Raw events are parsed by the two-pass algorithm described in
\S\ref{sec:pid}. The output is a set of windowed edge comma-separated value
(CSV) files
(\texttt{edges\_\{ts\}.csv}), each recording aggregated flow counts between
named service pairs over a configurable time window (default: 60 seconds).

\textbf{Phase 3: Spectral clustering.}
Windowed edge files are aggregated into a weighted undirected graph. Spectral
embedding followed by $k$-means and KL refinement assigns each service to a
VM partition (\S\ref{sec:cluster}).

\textbf{Phase 4: ROI scoring and plan generation.}
Given the VM partition and the dependency graph, the ROI scorer computes a
per-service migration priority score and emits a phased migration plan ranked
by ROI (\S\ref{sec:roi}).

\section{Implementation}

\subsection{PID-Port Correlation}
\label{sec:pid}

When microservices run as identically-named processes (e.g., all as
\texttt{python3} on the same host), eBPF events carry no usable service
identity in the \texttt{comm} or \texttt{service\_id} fields. All 20 services
in our testbed exhibited this pattern.

We solve the attribution problem with a two-pass algorithm over the raw event
stream:

\textbf{Pass 1: Build TGID to port map.}
For every accepted connection on a known service port, we record a vote for
that port against the accepting TGID. After processing all events, each TGID
is assigned to the port it accepted on most often.

\textbf{Pass 2: Attribute outbound calls.}
For every \texttt{dir=0} (connect) event to a known service port $d$, we look
up the originating TGID in the map from Pass~1 to obtain the source service
name, and emit the edge $\text{port}(\textit{tgid}) \to d$.

In our 20-service testbed, Pass~1 recovers all 20 TGIDs within the first 200
accept events, and attribution stays stable for the rest of the capture. No
service configuration files or process labels are needed. Because we authored
the testbed, the correct mapping was known in advance, so this establishes that
the attribution mechanism works rather than that it discovers an unknown
topology.

\subsection{Spectral Clustering with KL Refinement}
\label{sec:cluster}

Given the weighted dependency graph $G = (V, E, w)$, we compute the
normalized Laplacian $\mathcal{L} = I - D^{-1/2}AD^{-1/2}$ and embed each
service into $\mathbb{R}^{16}$ using the first 16 non-trivial eigenvectors of
$\mathcal{L}$~\cite{ng2001spectral}. $k$-means is then applied to this
embedding to produce an initial $k$-partition. With $|V| = 20$, a
16-dimensional embedding is only a modest reduction, so the embedding step
contributes less here than it would on a larger graph; most of the partition
quality comes from the refinement pass below.

Standard spectral clustering minimizes the ratio cut in the embedding space
rather than directly minimizing inter-partition edge weight. We add a
Kernighan-Lin (KL) refinement pass~\cite{kernighan1970efficient} that operates
on the original graph:

\begin{enumerate}
\item For each pair of clusters $(C_i, C_j)$, compute the \emph{D-value} for
      each node $v$:
      \[D(v) = w(v, \bar{C}_v) - w(v, C_v \setminus \{v\})\]
      where $C_v$ is the current cluster of $v$ and $\bar{C}_v$ is the other
      cluster in the pair.
\item Iteratively select the swap $(a \in C_i, b \in C_j)$ maximizing
      $D(a) + D(b) - 2w(a,b)$, accumulating swaps without applying them.
\item Apply the prefix of swaps with maximum cumulative gain.
\item Repeat until no improving swap exists or a pass limit is reached.
\end{enumerate}

KL refinement directly minimizes the inter-partition edge weight (cut\%),
making it better suited to our objective than the spectral embedding alone.
In our evaluation the target topology is two VMs: all 20 services begin
co-located on the source VM, and the partition assigns each service either to
remain there or to move to the second VM, so the clustering and refinement
operate on a two-way cut ($k=2$) throughout.

\subsection{ROI-Ranked Migration Ordering}
\label{sec:roi}

The partition from \S\ref{sec:cluster} determines \emph{which} VM each service
should occupy; the ordering problem determines \emph{when} to move each service
to minimize cross-VM traffic during the migration window.

We define the ROI score for service $S$ moving from $V_\text{src}$ to
$V_\text{dst}$ as:
\begin{equation}
  \text{ROI}(S) =
    \frac{\text{benefit}(S) - \text{cost}(S)}
         {\text{depth}(S) \times \text{variance}(S)}
  \label{eq:roi}
\end{equation}
where:
\begin{itemize}
  \item $\text{benefit}(S)$: total flows between $S$ and services already
        assigned to $V_\text{dst}$ (these become intra-VM after the move).
  \item $\text{cost}(S)$: total flows between $S$ and services remaining on
        $V_\text{src}$ (these become cross-VM after the move).
  \item $\text{depth}(S)$: in-degree of $S$ in the dependency graph (number
        of distinct callers); high depth indicates greater blast radius if
        the move disrupts service availability.
  \item $\text{variance}(S)$: standard deviation of $S$'s total flow volume
        across time windows; high variance indicates unpredictable load and
        higher migration risk.
\end{itemize}

Services are migrated in descending ROI order. High-ROI services have large
benefit, small cost, few callers, and stable load, so they are safe to move
early. Low-ROI services (hubs, high-variance services) move last, once their
dependencies are already on the target VM.

\section{Evaluation}
\label{sec:eval}

\subsection{Experimental Setup}

We deploy a 20-service microservice stack on two DigitalOcean droplets
(2~vCPU / 4~GB RAM each). All services run as Python 3 processes on ports
8001--8020, representing security, analytics, business, and infrastructure
tiers. Services communicate via the Hypertext Transfer Protocol (HTTP); the
gateway service (port 8020) acts as
the primary entry point for external load.

For dependency capture, we attach the nettrace eBPF daemon and drive load
against each service's \texttt{/workflow} endpoint, which triggers realistic
fan-out calls to downstream services. We capture 13,615 events over three
minutes, yielding 2,400 attributed flow records across 4 time windows.

For overhead measurement, we use a single-stack A/B design: 5 minutes of
warmup, followed by a 3-minute baseline phase (collector off) and a 3-minute
measurement phase (collector on, no restart). Load is generated by
\texttt{wrk2}~\cite{wrk2} at a fixed rate of 200 requests per second (RPS)
against the MoveGroups
application programming interface (API) stack (nginx + Gunicorn + PostgreSQL +
Redis).

We measure overhead against this stack rather than against the Python testbed
because it is closer to a production deployment in its process model and
connection behavior, and we wanted the overhead number to mean something
outside the testbed. The tradeoff is that the two workloads differ in
connection establishment rate, which is precisely the quantity eBPF collection
cost scales with. The figures in \S\ref{sec:overhead} therefore characterize
this stack, and should not be read as the overhead the testbed of
\S\ref{sec:eval} would have incurred.

\subsection{Dependency Graph Discovery}

\begin{table}[t]
\centering
\caption{Top dependency edges discovered by PID-port correlation}
\label{tab:edges}
\begin{tabular}{llr}
\toprule
Source & Destination & Flows \\
\midrule
search-service       & catalog-service       & 86 \\
auth-service         & fraud-detection       & 85 \\
catalog-service      & cache-service         & 85 \\
catalog-service      & image-service         & 85 \\
catalog-service      & inventory-service     & 85 \\
fraud-detection      & analytics-engine      & 85 \\
gateway-service      & auth-service          & 85 \\
shipping-service     & order-service         & 85 \\
cart-service         & pricing-service       & 66 \\
gateway-service      & cart-service          & 17 \\
\bottomrule
\end{tabular}
\end{table}

Pass~1 of the PID-port correlation algorithm recovers all 20 service TGIDs
within the first captured accept events. Table~\ref{tab:edges}
shows the 10 highest-flow edges; the full graph contains 32 edges spanning
all 20 services. Notable structural features: \texttt{catalog-service} is the
highest in-degree node (6 distinct callers), \texttt{audit-service} serves as
a logging hub for operational services, and \texttt{analytics-engine} is the
analytics sink for the fraud, pricing, and recommendation tiers.

These relationships were reconstructed automatically from three minutes of
runtime observation, without reading any service configuration. Two caveats
apply. We wrote the testbed, so the call graph was documented by construction;
the result shows the pipeline recovers a known graph from kernel events, not
that it surfaces relationships nobody knew. And the flow counts in
Table~\ref{tab:edges} are near-uniform (85--86 on most edges) because load was
driven at a fixed rate against each service's \texttt{/workflow} endpoint.
Real dependency graphs are heavily skewed, with long tails and code paths that
fire rarely or only under specific conditions; we have no evidence about
coverage or attribution accuracy in that regime.

\subsection{eBPF Collection Overhead}
\label{sec:overhead}

\begin{table}[t]
\centering
\caption{eBPF collection overhead (controlled A/B, single stack)}
\label{tab:overhead}
\begin{tabular}{lrrr}
\toprule
Metric        & Baseline (OFF) & Collector ON & Delta \\
\midrule
RPS           & 199.6          & 190.7        & $-4.4\%$ \\
p50 latency   & 112~ms         & 542~ms       & $+383\%$ \\
p99 latency   & 688~ms         & 7{,}910~ms   & $+1{,}050\%$ \\
Socket errors & 0              & 0            & --- \\
Non-2xx       & 0              & 0            & --- \\
\bottomrule
\end{tabular}
\end{table}

Table~\ref{tab:overhead} shows the overhead results. Throughput impact is
negligible ($-4.4\%$). Tail latency, however, increases substantially at
saturation: p50 rises by $383\%$ and p99 by $1{,}050\%$. We attribute this
to CPU contention between the eBPF ring-buffer drain and the Gunicorn worker
threads under near-saturation load (200~RPS against a 2-vCPU host). If that
mechanism is the right one, the effect should diminish where the host has
spare CPU headroom. We did not measure any lower load level, so this remains a
hypothesis that our data does not test, and the numbers above should be read as
characterizing the saturated case only.

eBPF collection is not free under load. Operators should run dependency captures
during off-peak periods or on dedicated sampling nodes rather than under
production saturation.

\subsection{Migration Ordering Comparison}

\begin{figure}[t]
\centering
\includegraphics[width=\columnwidth]{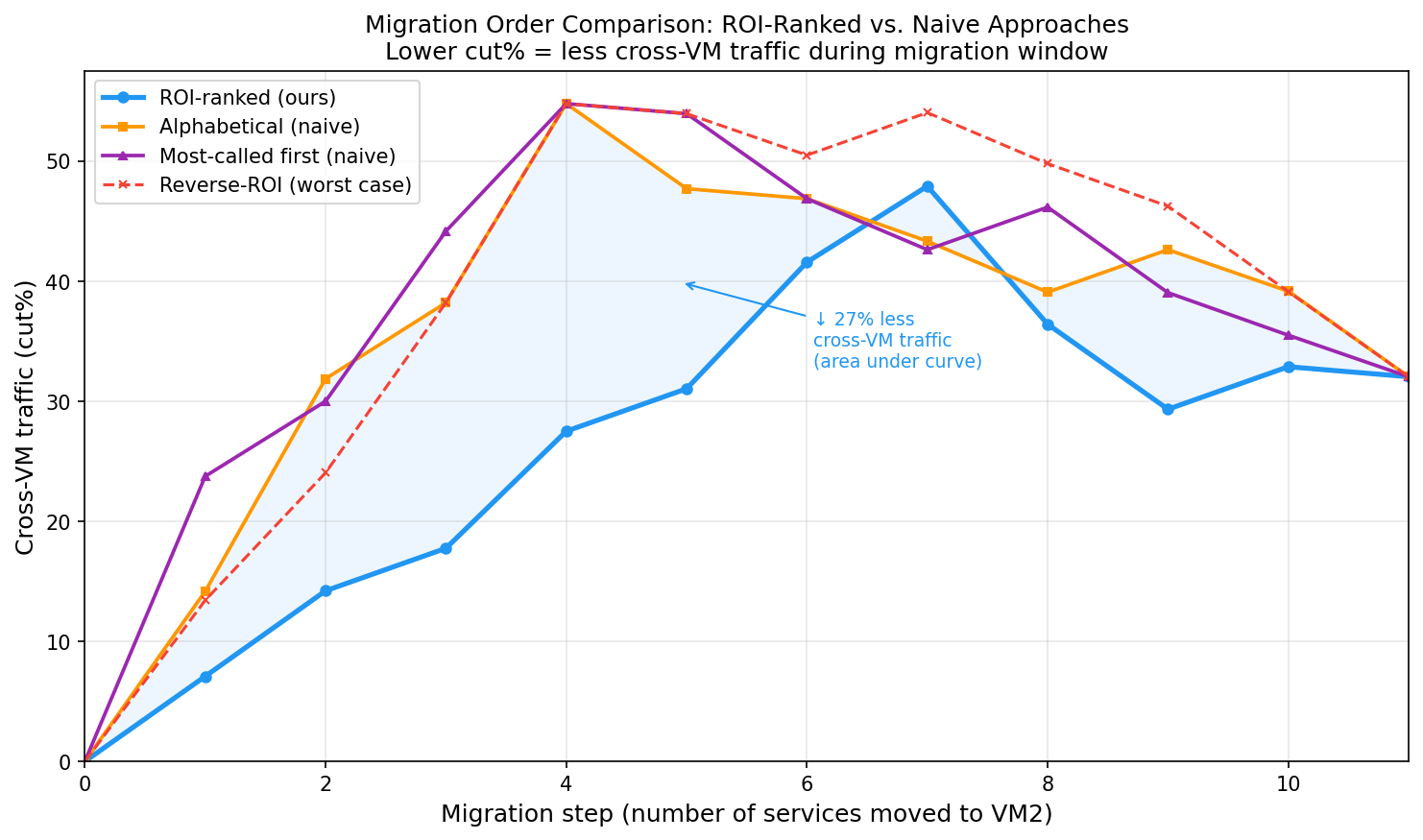}
\caption{Simulated cross-VM traffic (cut\%) at each migration step for our
ROI-ranked order versus three naive baselines, over the dependency graph
discovered from our 20-service testbed. Lower area under curve = less cross-VM
traffic exposure during the migration window. In simulation, our approach
reduces cumulative exposure by 27\% versus alphabetical ordering and 30\%
versus most-called-first; no services were actually migrated.}
\label{fig:migration}
\end{figure}

Fig.~\ref{fig:migration} shows the cut\% trajectory for four migration
orderings applied to our 20-service dependency graph. We define cut\% as the
fraction of total observed flow volume that crosses VM boundaries at each step.
The orderings are evaluated by simulating each migration sequence over the
discovered graph; no services were actually moved, and the simulation assumes
the graph is static across the migration window.

We compare:
\begin{itemize}
  \item \textbf{ROI-ranked (ours)}: services ordered by Eq.~(\ref{eq:roi}).
  \item \textbf{Alphabetical}: lexicographic order by service name, a common
        default in runbook-driven migrations. Because service names carry no
        information about the communication graph, this baseline serves as a
        proxy for an arbitrary (random) ordering, with the advantage of being
        deterministic and reproducible.
  \item \textbf{Most-called first}: services ordered by in-degree descending,
        representing an intuitive but incorrect heuristic (move busy services
        to their new home early).
  \item \textbf{Reverse-ROI}: deliberately worst case.
\end{itemize}

\begin{table}[t]
\centering
\caption{Area under cut\% curve (lower = better)}
\label{tab:auc}
\begin{tabular}{lr}
\toprule
Ordering & AUC \\
\midrule
ROI-ranked (ours)      & 301.7 \\
Alphabetical           & 413.9 \\
Most-called first      & 432.9 \\
Reverse-ROI (worst)    & 440.1 \\
\bottomrule
\end{tabular}
\end{table}

Table~\ref{tab:auc} reports the area under the cut\% curve (AUC), so that lower
values indicate less cumulative cross-VM traffic exposure. Our approach
achieves an AUC of 301.7, representing a $27.1\%$ reduction versus the
alphabetical proxy for arbitrary ordering and a $30.3\%$ reduction versus
most-called-first. These are single numbers from
one graph, one target topology ($k=2$), and one migration granularity; we did
not vary any of these, so the magnitude should not be expected to transfer. The improvement is
concentrated in the first half of the migration window (steps 1--6), where our
approach moves leaf services with zero cross-VM cost first, while naive
orderings immediately expose high-traffic hub dependencies.

The first three services in our ranked order (\texttt{image-service},
\texttt{search-service}, \texttt{metrics-collector}) have zero callers on
the source VM; moving them creates no new cross-VM edges. Naive orderings
move \texttt{analytics-engine} or \texttt{audit-service} early due to
alphabetical precedence, immediately creating 4--6 cross-VM edges that persist
for the remainder of the migration.

\section{Discussion}

\textbf{Generalizability.}
The PID-port correlation algorithm assumes each service consistently listens on
a stable port. This holds for the vast majority of containerized and VM-hosted
deployments. It would not hold for services using dynamic port binding or
multi-port services where the same TGID accepts on multiple well-known ports;
in those cases, the voting heuristic would need augmentation with container
metadata.

\textbf{Capture duration.}
Three minutes of targeted load was sufficient to discover 32 dependency edges
in our testbed. In production environments with more complex call graphs or
infrequent code paths, longer captures or endpoint-directed synthetic load may
be necessary to achieve complete coverage.

\textbf{Tail latency under saturation.}
The p99 latency spike observed in \S\ref{sec:overhead} is a real operational
concern. We recommend running eBPF captures during scheduled maintenance
windows or using sampling (capture 1-in-$N$ connections) to reduce ring-buffer
pressure under load.

\textbf{Dynamic graphs.}
Microservice dependency graphs change as services are updated or traffic
patterns shift. The pipeline supports incremental recapture; re-running
Phases 1--2 on a new capture and comparing the resulting graph against the
previous partition can detect drift and trigger replanning.

\section{Limitations}
\label{sec:limits}

We report a working end-to-end pipeline evaluated on one synthetic testbed. The
following constraints bound what that evaluation supports.

\textbf{Authored ground truth.} We wrote the 20-service testbed, chose its
topology, and generated its load. Recovering that topology validates the
attribution and clustering machinery; it does not demonstrate discovery of
unknown dependencies, and it does not exercise the conditions that make real
dependency discovery hard, namely skewed traffic, rare code paths, and services
whose behavior varies with input.

\textbf{Simulated migration outcome.} The 27\% figure comes from simulating
orderings over the discovered graph, not from performing migrations. The
simulation holds the graph fixed across the window and ignores migration
duration, partial availability, and retry traffic during cutover. We have not
varied the topology, the number of target VMs, or the graph size, so we cannot
say how sensitive the result is to any of them.

\textbf{Overhead measured on a different stack and one load level.} Overhead was
measured against the MoveGroups API stack rather than the testbed, at a single
rate of 200~RPS on a 2-vCPU host. Connection establishment rate differs between
the two workloads, and we have no measurements at lower utilization, so the
overhead profile across the operating range is unknown.

\textbf{Scale.} Twenty services across two VMs is small. Spectral partitioning
and KL refinement are both well characterized at larger scales. The parts that
worry us are the ones we measured: eBPF event volume, the ring-buffer pressure
behind the tail latency effect, and whether PID-port voting stays stable when
hundreds of processes contend for the same well-known ports.

\textbf{No comparison against existing dependency sources.} We argue that our
approach avoids the operational cost of a service mesh, but we never put the
two side by side. Running Istio or Jaeger against the same workload and diffing
the resulting graphs against ours is the obvious way to show that kernel-level
observation misses nothing important. It is also the experiment we would run
first if we continued this work.

\section{Conclusion}

We built an end-to-end system for dependency-aware microservice migration
planning using eBPF network tracing. Its three core pieces are PID-port
correlation for service attribution, spectral clustering with KL refinement
for VM partitioning, and ROI-ranked migration ordering. Together they replace
ad-hoc migration sequencing with an approach grounded in live runtime
behavior.

On a 20-service testbed, the pipeline recovers the full inter-service
dependency graph from three minutes of traffic observation, and in simulation
its ROI-ranked ordering reduces cumulative cross-VM traffic exposure by 27\%
relative to alphabetical ordering. Collection overhead is asymmetric: at
near-saturation load on a 2-vCPU host, throughput dropped 4.4\% while p99
latency rose by an order of magnitude, which is a real constraint on when
captures can be run.

The pipeline needs no code changes, no service mesh, and no help from service
developers. Any operator with root access to the host can attach the eBPF probe
and get a dependency-aware migration plan in under 10 minutes. Whether the
planning benefit survives contact with a production dependency graph, and
whether the overhead profile improves with CPU headroom, are the two questions
we would want answered next.

\section*{Acknowledgment}

AI-based writing assistance was used in drafting and editing the text of this
paper. The system design, implementation, experiments, and all reported
measurements are the authors' own work; every reference was verified against
its source by the authors prior to submission.


\end{document}